\documentclass[floats,floatfix,showpacs,amssymb,prd,
superscriptaddress,nofootinbib]{revtex4-1}

\usepackage{amssymb,amsmath,verbatim,mathtools,needspace,enumitem,etoolbox,graphicx,physics,microtype,afterpage,xspace,tabularx,lmodern,multirow}
\usepackage{gensymb}

\usepackage[dvipsnames, usenames]{xcolor}
\definecolor{linkcolor}{rgb}{0.0,0.3,0.5}
\usepackage[unicode, colorlinks=true, linkcolor=linkcolor, citecolor=linkcolor, filecolor=linkcolor, urlcolor=linkcolor, linktocpage, breaklinks]{hyperref}
\usepackage{dsfont}
\usepackage[all]{hypcap}
\usepackage[T1]{fontenc}
\usepackage[utf8]{inputenc}
\usepackage[usenames,dvipsnames]{xcolor}
\hypersetup{colorlinks=true,citecolor=romared,linkcolor=romared,urlcolor=romared}

\definecolor{romared}{RGB}{142,0,28}

\newcommand{\be}{\begin{equation}}
	\newcommand{\ee}{\end{equation}}

\def\be{\begin{equation}}
	\def\ee{\end{equation}}
\newcommand{\beq}{\begin{eqnarray}}
	\newcommand{\eeq}{\end{eqnarray}}

\usepackage{makecell}
\usepackage{soul}

\begin{document}
	
\title{$sl(2,\mathbb{C})\times D$ symmetry and conformal primary basis for massless fields}

\author{Yuan Chen}
\email{yuanchen@mail.bnu.edu.cn}
\affiliation{Department of Physics, Beijing Normal University, Beijing 100875, China}
\author{Mingfeng Li}
\email{mingfeng.li@mail.bnu.edu.cn}
\affiliation{Department of Physics, Beijing Normal University, Beijing 100875, China}
\author{Kai Shi}
\email{kaishi@mail.bnu.edu.cn}
\affiliation{Department of Physics, Beijing Normal University, Beijing 100875, China}
\author{Hongbao Zhang}
\email{hongbaozhang@bnu.edu.cn}
\affiliation{Department of Physics, Beijing Normal University, Beijing 100875, China}
\affiliation{Key Laboratory of Multiscale Spin Physics, Ministry of Education, Beijing Normal University, Beijing 100875, China}
\author{Jingchao Zhang}
\email{jczhang@mail.bnu.edu.cn}
\affiliation{Department of Physics, Beijing Normal University, Beijing 100875, China}
\affiliation{Leinweber Center for Theoretical Physics, University of Michigan, Ann Arbor, MI 48109, USA}


\begin{abstract}
Alternative to the embedding formalism, we provide a group theoretic approach to the conformal primary basis for the massless field with arbitrary helicity. To this end, we first point out that $sl(2,\mathbb{C})$ isometry gets enhanced to $sl(2,\mathbb{C})\times D$ symmetry for the solution space of the massless field with $D$ the bulk dilatation. Then associated with $sl(2,\mathbb{C})\times D$ symmetry, we introduce the novel quadratic Casimirs and relevant tensor/spinor fields to derive two explicit constraints on the bulk dilatation and $sl(2,\mathbb{C})$ Casimirs. With this, we further argue that the candidate conformal primary basis can be constructed out of the infinite tower of the descendants of the left and right highest (lowest) conformal primary wavefunction of $sl(2,\mathbb{C})$ Lie algebra, and the corresponding celestial conformal weights are determined by the bulk scaling dimension through solving out the exact on-shell conformal primary wavefunctions, where on top of the two kinds of familiar-looking on-shell conformal primary wavefunctions, we also obtain another set of independent on-shell conformal primary wavefunctions for the massless field with helicity $|s|\ge 1$. In passing, we also develop the relationship between the 4D Lorentz Lie algebra and 2D conformal Lie algebra from scratch, and present an explicit derivation for the two important properties associated with the conformal primary wavefunctions. 
\end{abstract}

\maketitle
\flushbottom

\section{Introduction}
Over the last few decades, holographic principle has been standing out as a guiding principle for us to formulate the quantum theory of gravity. AdS/CFT correspondence, as one explicit implementation of such a principle, states that the quantum gravity in an asymptotically Anti-de Sitter spacetime is encoded fully by the boundary conformal field theory (CFT). On the other hand, by holography, the only observable in an asymptotically flat spacetime is the scattering amplitude. However, the scattering amplitude is expressed conventionally in the momentum representation, which manifests the translation symmetry but obscures the holographic nature. Given this, a new representation in terms of the so-called conformal primary basis has been constructed via the embedding formalism in CFT, whereby the scattering amplitude in the $d$-dimensional flat spacetime admits a natural holographic interpretation of the conformal correlator in the $(d-2)$-dimensional celestial sphere\cite{BS,CFS,PS,PSS,ST,FT,DPS,LZ1,LZ2,IM}. Such a holographic reformulation of the scattering process in terms of the so-called celestial amplitude further motivates a recently conjectured duality, called celestial holography, which proposes that the bulk scattering in the flat spacetime can be dual to a CFT on the celestial sphere\cite{Strominger,Pasterski1,Raclariu,Pasterski2,PPR}. 

No matter whether celestial holography turns out to be valid or not, such an alternative formulation of the scattering process has already shed new light on our understanding of scattering amplitude, where the conformal primary basis, as the building block of the whole reformulation, plays a vital role as it should be the case. In particular, for the case of massless particles, the conformal primary basis in the new representation turns out to be  related to the familiar plane wave basis in the momentum representation by a Mellin transformation or further followed by a shadow transformation. The main purpose of this paper is to offer a group theoretic understanding of the conformal primary basis for the massless particles in the $4$-dimensional Minkowski spacetime, which is supposed to serve as an alternative perspective to the aforementioned embedding formalism. To this end, we shall first develop the relationship between 4D Lorentz Lie algebra and 2D conformal Lie algebra from scratch and derive the two important properties associated with the conformal primary wavefunctions in the subsequent section. Then in Section \ref{group}, with the observation of the bulk dilatation $D$ as an emergent symmetry for massless particles, we will argue that the $sl(2,\mathbb{C})\times D$ symmetry dictated conformal primary basis can serve as  a candidate basis for the massless particle representation of the Poincar\'e Lie algebra, which is further substantiated by an explicit derivation. In Section \ref{correspondence}, we build the specific correspondence between the 4D bulk scaling dimension and 2D celestial conformal weights for all the on-shell conformal primary wavefunctions. We shall conclude our paper with some discussions in the final section.

Notation and conventions follow Chapter $13$ of \cite{Wald}, where lower and upper Latin indices denote Penrose's abstract notations for tensors and spinors, respectively while lower and upper Greek indices represent the corresponding components. In addition, where no confusion arises, lower Greek indices also denote the concrete indices, running from $0$ to $d-1$ while  the intermediate lower Latin letters $(i,j,k)$ denote the spatial components or concrete indices, running from $1$ to $d-1$. In particular, the signature is $(+,-,-,-)$, and $\epsilon_{0123}=1$. Furthermore, a spinor is raised and lowered as $\phi^A=\epsilon^{AB}\phi_B,\phi_B=\epsilon_{AB}\phi^A$
with $\epsilon_{AC}\epsilon^{BC}=\epsilon_A{}^B=-\epsilon^B{}_A=\delta_A{}^B$.

\section{4D Lorentz group, 2D conformal group, and  conformal primary wavefunctions}

In what follows, we shall focus exclusively onto the $4$-dimensional Minkowski spacetime, where we can take advantage of the spinor machinary to develop the relationship between 4D Lorentz group and 2D conformal group.

To proceed, we like to take the canonical choice of spinor basis $o^A, \iota^A$ such that $o_A\iota^A=1$, with the dual basis given by $-\iota_A, o_A$. Furthermore,  we have $\epsilon^{AB}=o^A\iota^B-\iota^Ao^B$ and $\epsilon_{AB}=o_A\iota_B-\iota_Ao_B$, which means 
\begin{equation}
	\epsilon^{\Sigma\Gamma}=\begin{pmatrix}0&1\\-1&0\end{pmatrix},\quad
	\epsilon_{\Sigma\Gamma}=\begin{pmatrix}0&1\\-1&0\end{pmatrix}.
\end{equation}
The identification between spinors in the above basis and tensors in the Lorentz coordinates is specified by the soldering form $\sigma^\mu{}_{\Sigma\Gamma'}=\frac{1}{\sqrt{2}}(\mathds{1},\sigma^i)$ with $\mathbf{\sigma}$ the standard Pauli matrices
\begin{equation}
	\sigma^1=\begin{pmatrix}0&1\\1&0\end{pmatrix},\quad \sigma^2=\begin{pmatrix}0&-i\\i&0\end{pmatrix},\quad \sigma^3=\begin{pmatrix}1&0\\0&-1\end{pmatrix}.
\end{equation}
Whence one also has $\sigma_\mu{}^{\Sigma'\Gamma}=\frac{1}{\sqrt{2}}(\mathds{1},\sigma^i)$. In particular, our Minkowski metric is related to its spinor representation as follows
\begin{equation}	\eta_{\mu\nu}=\sigma_\mu{}^{\Sigma'\Gamma}\sigma_\nu{}^{\Omega'\Xi}\epsilon_{\Sigma'\Omega'}\epsilon_{\Gamma\Xi},
\end{equation}
where the complex conjugation is indicated implicitly by the primed indices. With this, the identification $SO(1,3)\simeq SL(2,\mathbb{C})/\mathbb{Z}_2$ can be expressed as follows 
\begin{equation}
	\Lambda^\mu{}_\nu x^\nu\sigma_\mu{}^{\Sigma'\Gamma}=x^\mu \bar{L}^{\Sigma'}{}_{\Omega'}\sigma_{\mu}{}^{\Omega'\Xi}L^\Gamma{}_\Xi,
\end{equation}
where $\Lambda\in SO(1,3)$ and $L\in SL(2,\mathbb{C})$, satisfying $L^\Sigma{}_\Gamma L^\Omega{}_\Xi\epsilon_{\Sigma\Omega}=\epsilon_{\Gamma\Xi}$. Accordingly, the corresponding Lie algebras can be related to each other as 
\begin{equation}
	\lambda^\mu{}_\nu x^\nu\sigma_\mu{}^{\Sigma'\Gamma}=x^\mu (\bar{l}^{\Sigma'}{}_{\Omega'}\sigma_{\mu}{}^{\Omega'\Gamma}+\sigma_\mu{}^{\Sigma'\Xi}l^\Gamma{}_\Xi).
\end{equation}
Note that the Pauli matrices serve naturally as the generators for $sl(2,\mathbb{C})$. A straightforward manipulation of the Pauli matrices leads to the following realization of $sl(2,\mathbb{C})$
\begin{eqnarray}
&&	\sigma^1\rightarrow -2\mathbf{K}_1,\quad \sigma^2\rightarrow 2\mathbf{K}_2,\quad \sigma^3\rightarrow -2\mathbf{K}_3, \nonumber \\ 
&& i\sigma^1\rightarrow -2\mathbf{J}_1, \quad i\sigma^2\rightarrow 2\mathbf{J}_2, \quad i\sigma^3\rightarrow -2\mathbf{J}_3 
\end{eqnarray}
with the Lorentz boosts and rotations defined as\footnote{Kindly please refer to Appendix \ref{A} for all the generators of the conformal algebra in the $d$-dimensional Minkowski spacetime.}
\begin{equation}
	\mathbf{K}_i=M_{i0}, \quad 
 \mathbf{J}_i=\frac{1}{2}\epsilon_{0ijk}M^{jk}.
\end{equation}
However, one can also reach the following two commutative  realizations of $sl(2,\mathbb{C})$ by the complexified Lorentz generators
\begin{eqnarray}\label{1}
	&& \sigma^1\rightarrow -\mathbf{K}_1+i\mathbf{J}_1,\quad    \sigma^2\rightarrow \mathbf{K}_2-i\mathbf{J}_2,\nonumber \\   
	&& \sigma^3\rightarrow -\mathbf{K}_3+i\mathbf{J}_3, \quad  i\sigma^1\rightarrow i(-\mathbf{K}_1+i\mathbf{J}_1),\nonumber \\   &&i\sigma^2\rightarrow i(\mathbf{K}_2-i\mathbf{J}_2),\quad i\sigma^3\rightarrow i(-\mathbf{K}_3+i\mathbf{J}_3) 
\end{eqnarray}
from
\begin{equation}
	\lambda^\mu{}_\nu x^\nu\sigma_\mu{}^{\Sigma'\Gamma}=x^\mu\sigma_\mu{}^{\Sigma'\Xi}l^\Gamma{}_\Xi,
\end{equation}
and 
\begin{eqnarray}\label{2}
	&&\overline{\sigma^1}\rightarrow -\mathbf{K}_1-i\mathbf{J}_1,\quad \overline{\sigma^2}\rightarrow \mathbf{K}_2+i\mathbf{J}_2,\nonumber\\
 &&\overline{\sigma^3}\rightarrow -\mathbf{K}_3-i\mathbf{J}_3, \quad \overline{i\sigma^1}\rightarrow -i(-\mathbf{K}_1-i\mathbf{J}_1),\nonumber\\ 
 &&\overline{i\sigma^2}\rightarrow -i(\mathbf{K}_2+i\mathbf{J}_2),\quad \overline{i\sigma^3}\rightarrow-i( -\mathbf{K}_3-i\mathbf{J}_3)
\end{eqnarray}
from 
\begin{equation}
	\lambda^\mu{}_\nu x^\nu\sigma_\mu{}^{\Sigma'\Gamma}=x^\mu \bar{l}^{\Sigma'}{}_{\Omega'}\sigma_{\mu}{}^{\Omega'\Gamma},
\end{equation}
where the overline indicates that the corresponding realization is a complex realization. 

On the other hand, $SL(2,\mathbb{C})$ can also be understood as the global conformal group on the celestial sphere. To this end, let $\lambda^\Sigma=(w,1)$, then its corresponding null vector $\lambda^{\Sigma}\lambda^{\Sigma'}$ is given by $q^\mu=(w\bar{w}+1, w+\bar{w},i(\bar{w}-w),w\bar{w}-1)$, whose spatial component can be geometrized as a point on a unit celestial sphere as $\mathbf{q}=(w\bar{w}+1)\hat{\mathbf{q}}$ by performing the stereographic projection from the north pole of the sphere to the complex plane with $w=\cot\frac{\theta}{2}e^{i\varphi}$. Similarly, with the choice of $\lambda^\Sigma=(1,\bar{w})$, the corresponding null vector is given by $q^\mu=(1+w\bar{w},w+\bar{w},i(\bar{w}-w),1-w\bar{w})$, which can be visualized as a point on the unit celestial sphere as $\mathbf{q}=(1+w\bar{w})\hat{\mathbf{q}}$ instead by performing the stereographic projection from the south pole to the complex plane with $w=\tan \frac{\theta}{2}e^{i\varphi}$. In what follows, we prefer to work exclusively with the first parametrization of our spinor as well as its corresponding null vector\footnote{As to the second parametrization with $\bar{w}$ replaced by $w$, kindly please refer to Appendix \ref{B} for the relevant results.}. 
Accordingly, $SL(2,\mathbb{C})$ acting on our spinor $\lambda^\Sigma$ will induce a global conformal transformation on the celestial sphere as follows
\begin{equation}\label{coordinatetrans}
	w'= \begin{pmatrix}a&b\\c&d\end{pmatrix} w=\frac{aw+b}{cw+d}, \quad \bar{w}'=\begin{pmatrix}\bar{a}&\bar{b}\\\bar{c}&\bar{d}\end{pmatrix} \bar{w}=\frac{\bar{a}\bar{w}+\bar{b}}{\bar{c}\bar{w}+\bar{d}},
\end{equation}
with $ad-bc=1$, where $w$ and $\bar{w}$ are assumed to be independent of each other. 
Whence it is not hard to show that $sl(2,\mathbb{C})$ can be realized respectively on $w$ and $\bar{w}$ space as follows
\begin{eqnarray}\label{3}
	&&l_{-1}=\begin{pmatrix}0&1\\0&0\end{pmatrix}=\frac{1}{2}(\sigma^1+i\sigma^2)\rightarrow T_{-1}, \nonumber \\ &&\bar{l}_{-1}=\begin{pmatrix}0&1\\0&0\end{pmatrix}=\frac{1}{2}(\overline{\sigma^1+i\sigma^2})\rightarrow \bar{T}_{-1},
	\nonumber\\
	&&l_1=\begin{pmatrix}0&0\\-1&0\end{pmatrix}=-\frac{1}{2}(\sigma^1-i\sigma^2)\rightarrow T_1, \nonumber \\ &&\bar{l}_1=\begin{pmatrix}0&0\\-1&0\end{pmatrix}=-\frac{1}{2}(\overline{\sigma^1-i\sigma^2})\rightarrow \bar{T}_1,\nonumber
	 \\
	&&l_0=\frac{1}{2}\begin{pmatrix}1&0\\0&-1\end{pmatrix}=\frac{1}{2}\sigma^3
	\rightarrow T_0, \nonumber \\
	 && \bar{l}_0=\frac{1}{2}\begin{pmatrix}1&0\\0&-1\end{pmatrix}=\frac{1}{2}\overline{\sigma^3}
	\rightarrow \bar{T}_0,
\end{eqnarray}
where the overline also indicates that the corresponding realization is a complex one, with $T_n=w^{n+1}\partial_w$ and $\bar{T}_n=\bar{w}^{n+1}\partial_{\bar{w}}$ the vector fields on the celestial sphere, satisfying the following commutation relations
\begin{eqnarray}
\left[T_n, T_m\right]&=&(m-n)T_{n+m},\nonumber\\ 
 \left[\bar{T}_n,\bar{T}_m\right]&=&(m-n)\bar{T}_{n+m}, \nonumber\\ 
 \left[T_n,\bar{T}_m\right]&=& 0.   
\end{eqnarray}

By inspection of Eq. (\ref{1}), Eq. (\ref{2}), and Eq. (\ref{3}), one can obtain the identification between the 4D Lorentz generators and 2D conformal generators
\begin{eqnarray}
	&&T_{-1}\simeq L_{-1}=\frac{1}{2}(-\mathbf{K}_1+\mathbf{J}_2+i(\mathbf{K}_2+\mathbf{J}_1)),\quad \nonumber\\
	&&\bar{T}_{-1}\simeq \bar{L}_{-1}=\frac{1}{2}(-\mathbf{K}_1+\mathbf{J}_2-i(\mathbf{K}_2+\mathbf{J}_1)),\nonumber\\
	&&T_1\simeq L_1=\frac{1}{2}(\mathbf{K}_1+\mathbf{J}_2+i(\mathbf{K}_2-\mathbf{J}_1)),\quad \nonumber \\
	&&\bar{T}_1\simeq \bar{L}_1=\frac{1}{2}(\mathbf{K}_1+\mathbf{J}_2-i(\mathbf{K}_2-\mathbf{J}_1)),\nonumber\\
	&&l_0\simeq	L_0=\frac{1}{2}(-\mathbf{K}_3+i \mathbf{J}_3),\quad \nonumber\\
	&& \bar{l}_0\simeq\bar{L}_0=\frac{1}{2}(-\mathbf{K}_3-i \mathbf{J}_3).
\end{eqnarray}
Accordingly, we have the following commutation relations 
\begin{eqnarray}
 \left[L_n,L_m\right]&=&(m-n)L_{m+n},\nonumber\\
 \left[\bar{L}_n, \bar{L}_m\right]&=&(m-n)\bar{L}_{m+n},\nonumber\\
 \left[L_n,\bar{L}_m\right]&=&0  
\end{eqnarray}
with $n,m=0,\pm1$. For later convenience, we would like to denote the Lie algebras out of $L_n$ and $\bar{L}_n$ as $sl(2,\mathbb{C})_L$ and $sl(2,\mathbb{C})_R$, respectively.

A wavefunction on our Minkowski spacetime and the celestial sphere is called the conformal primary wavefunction with the $SL(2,\mathbb{C})$ conformal dimension $\Delta$ and spin $J$ if 
\begin{eqnarray}\label{celestial}
&&\mathcal{O}(x'^\mu=\Lambda^\mu{}_\nu x^\nu;w'=\frac{aw+b}{cw+d},\bar{w}'=\frac{\bar{a}\bar{w}+\bar{b}}{\bar{c}\bar{w}+\bar{d}})\nonumber\\
&&=|\frac{\partial w'}{\partial w}|^{-\frac{\Delta+J}{2}}|\frac{\partial\bar{w}'}{\partial\bar{w}}|^{-\frac{\Delta-J}{2}}D(\Lambda)\mathcal{O}(x;w,\bar{w}),    
\end{eqnarray}
where the representation of Lorentz group $D(\Lambda)$ is determined by the spinor and tensor indices of $\mathcal{O}$ as usual with $|\frac{\partial w'}{\partial w}|=\frac{1}{(cw+d)^2}$ and $|\frac{\partial\bar{w}'}{\partial \bar{w}}|=\frac{1}{(\bar{c}\bar{w}+\bar{d})^2}$. Whence we have 
\begin{eqnarray}
 &&D(\Lambda)^{-1}\mathcal{O}(\Lambda x;w',\bar{w}')-\mathcal{O}(x;w',\bar{w}')\nonumber
 \\&&=|\frac{\partial w'}{\partial w}|^{-h}|\frac{\partial\bar{w}'}{\partial\bar{w}}|^{-\bar{h}}\mathcal{O}(x;w,\bar{w})-\mathcal{O}(x;w',\bar{w}'),  
\end{eqnarray}
where the $SL(2,\mathbb{C})$ conformal weights are given by $(h,\bar{h})=\frac{1}{2}(\Delta+J,\Delta-J)$. This implies 
\begin{eqnarray}\label{reptrans}
 && \mathcal{L}_{L_n}\mathcal{O}=-\mathcal{L}_{T_n}\mathcal{O}=-(w^{n+1}\partial_w+h(n+1)w^n)\mathcal{O},\nonumber\\
	&&\mathcal{L}_{\bar{L}_n}\mathcal{O}=-\mathcal{L}_{\bar{T}_n}\mathcal{O}=-(\bar{w}^{n+1}\partial_{\bar{w}}+\bar{h}(n+1)\bar{w}^n)\mathcal{O}  \nonumber\\
\end{eqnarray}
for $n=0,\pm1$, which amounts to saying that the Lie derivative of the Lorentz generators acting on the conformal primary wavefunction can be understood as the Lie derivative of the corresponding $SL(2,\mathbb{C})$ generators acting on it with an additional minus sign. This is essentially the underlying reason for the definition of the  conformal primary wavefunction through Eq. (\ref{celestial}). For our purpose, we would like to list the celestial conformal weights and the bulk scaling dimension for a few important conformal primary wavefunctions in Table \ref{list1}. Furthermore, if a conformal primary wave function is on-shell, namely satisfies the equation of motion dictated by the unitary representation of the Poincar\'e group, one can define an operator on the celestial sphere associated with it as follows
\begin{equation}
	\hat{\mathcal{O}}(w',\bar{w}')=(\hat{\Phi}(x'), \mathcal{O}(x';w',\bar{w}'))_{\Sigma'},
\end{equation}
where $\hat{\Phi}(x')$ is the corresponding bulk quantum field and the Klein-Gordon inner product $(,)_{\Sigma'}$ evaluated on a Cauchy surface $\Sigma'$ is nevertheless independent of the choice of $\Sigma'$. The Lorentz covariance of the Klein-Gordon inner product implies
\begin{eqnarray}
	&&\hat{\mathcal{O}}(w',\bar{w}')\nonumber\\
& &=(D(\Lambda)^{-1}\hat{\Phi}(x'),D(\Lambda)^{-1}\mathcal{O}(x';w',\bar{w}'))_\Sigma\nonumber\\
&	&=(U(\Lambda)\hat{\Phi}(x)U(\Lambda)^{-1},|\frac{\partial w'}{\partial w}|^{-h}|\frac{\partial\bar{w}'}{\partial\bar{w}}|^{-\bar{h}}\mathcal{O}(x;w,\bar{w}))_\Sigma\nonumber\\
	&&=|\frac{\partial w'}{\partial w}|^{-h}|\frac{\partial\bar{w}'}{\partial\bar{w}}|^{-\bar{h}}U(\Lambda)\hat{\mathcal{O}}(w,\bar{w})U(\Lambda)^{-1},
\end{eqnarray}
where $U(\Lambda)$ corresponds to the unitary representation of the Lorentz group in the Fock space. It is noteworthy that due to its presence, the celestial operator $\hat{\mathcal{O}}$ does not transform under the global conformal transformation as the ordinary conformal primary operators. But nevertheless, due to the Lorentz invariance of both the vacuum and S-matrix, i.e., $U(\Lambda)|0\rangle=|0\rangle$ and $U(\Lambda)^{-1}SU(\Lambda)=S$, we have
\begin{widetext}
\begin{equation}
	 \langle 0 |\hat{\mathcal{O}}_i(w'_i,\bar{w}'_i) S\hat{\mathcal{O}}_j(w'_j,\bar{w}'_j)|0\rangle
	=|\frac{\partial w'_i}{\partial w_i}|^{-h_i}|\frac{\partial\bar{w}'_i}{\partial\bar{w}_i}|^{-\bar{h}_i}|\frac{\partial w'_j}{\partial w_j}|^{-h_j}|\frac{\partial\bar{w}'_j}{\partial\bar{w}_j}|^{-\bar{h}_j} \langle 0 |\hat{\mathcal{O}}_i(w_i,\bar{w}_i) S\hat{\mathcal{O}}_j(w_j,\bar{w}_j)|0\rangle,
\end{equation}
\end{widetext}
which tells us that the celestial amplitude behaves like the conformal correlator on the celestial sphere.

\begin{table}
	\begin{center}
 \begin{ruledtabular}
		\begin{tabular}{c|c|c|c|c|c}
			& $h$ & $\bar{h}$& $\Delta$ & $J$ & $\mathcal{D}$\\\hline
			$\lambda^\Sigma$  & $-\frac{1}{2}$& $0$& $-\frac{1}{2}$&$-\frac{1}{2}$ & $-\frac{1}{2}$\\
			$\bar{\lambda}^{\Sigma'}$&$0$ & $-\frac{1}{2}$ & $-\frac{1}{2}$&$\frac{1}{2}$ & $-\frac{1}{2}$\\
			$q^\mu$& $-\frac{1}{2}$& $-\frac{1}{2}$   &$-1$& $0$ & $-1$\\
			$D^\mu$ & $0$ & $0$ &$0$ & $0$ & $0$\\
			$\epsilon_{\Sigma\Gamma}$ &$0$& $0$ & $0$ & $0$ & $1$\\
			$\eta_{\mu\nu}$ & $0$&$0$ &$0$ & $0$ & $2$\\
			$ \frac{1}{ q\cdot x}$&$\frac{1}{2}$ & $\frac{1}{2}$ & $1$ & $0$ & $-1 $\\
			$\frac{\lambda^\Sigma}{\sqrt{q\cdot x}}$ & $-\frac{1}{4}$ & $\frac{1}{4}$ & $0$ & $-\frac{1}{2}$ & $-1$ \\
			$\frac{D^{\Sigma\Sigma'}\bar{\lambda}_{\Sigma'}}{\sqrt{q\cdot x}}$ & $\frac{1}{4}$ & $-\frac{1}{4}$ & $0$ & $\frac{1}{2}$ & $0$\\
		\end{tabular}
		\caption{The celestial conformal weights (conformal dimension and spin) and the bulk scaling dimension for some conformal primary wavefunctions.}
		\label{list1}	
  \end{ruledtabular}
	\end{center}		
\end{table}


\section{$sl(2,\mathbb{C})\times D$ symmetry and the candidate basis for the massless particle representation of the Poincar\'e group}\label{group}
The unitary representation of the Poincar\'e group for massless particles is usually expressed in terms of the simultaneous eigenvectors of the commuting spatial $3$-momentum or the commuting $4$-momentum with one on-shell condition $P_\mu P^\mu=0$. But note that $L^2, L_0$ and $\bar{L}^2,\bar{L}_0$ commuting with one another, where 
\begin{eqnarray}
	L^2&=&L_0^2-\frac{1}{2}(L_{-1}L_1+L_1L_{-1}),\nonumber\\
 \bar{L}^2&=&\bar{L}_0^2-\frac{1}{2}(\bar{L}_{-1}\bar{L}_1+\bar{L}_1\bar{L}_{-1})
\end{eqnarray}
are the corresponding Casimir operators for $sl(2,\mathbb{C})_L$  and $sl(2,\mathbb{C})_R$ Lie algebras, respectively. So it is reasonable to expect the simultaneous eigenvectors of the above $4$ operators could constitute the candidate basis as well for the massless particle representation of the Poincar\'e group. But as alluded above, the number of degrees of freedom for the $4$-momentum is not $4$ but $3$ due to the on-shell condition. So there may exist one similar constraint on $L^2, L_0,\bar{L}^2,\bar{L}_0$. As we shall show in this section, this is the case indeed, where it turns out that the bulk dilation operator come to play a crucial role. To be more specific, first
we note that not only does the bulk dilatation vector field $D$ together with the Killing vector fields form a closed Lie algebra, but also commutes with the two Poincar\'e Casimir operators $P^2=P_\mu P^\mu$ and $W^2=W_\mu W^\mu$ with the Pauli-Lubanski spin operator defined as $W_\mu=-\frac{1}{2}\epsilon_{\mu\nu\rho\sigma}P^\nu M^{\rho\sigma}$ for the massless particle representation of the Poincar\'e Lie algebra, where $W_\mu=isP_\mu$ with $s$ the helicity of the massless particle\footnote{For the massless particle representation of the Poincar\'e group, we have $[D, P^2]=-2P^2=0$. In addition, $[D,P_\mu]=-P_\mu$ together with $[D, W_\mu]=-W_\mu$ implies that $D$ commutes with the helicity $s$.}. This amounts to saying that the bulk dilatation emerges as an additional symmetry of the solution space of the equations of motion dictated by the massless particle representation of the Poincar\'e group. In addition, $D$ commutes with the Lorentz boosts and rotations, thus also commutes with $L^2, L_0,\bar{L}^2,\bar{L}_0$. As a result, associated with $sl(2,\mathbb{C})\times D$ symmetry, the candidate basis is formed virtually by the simultaneous eigenvectors of $5$ operators $D, L^2, L_0,\bar{L}^2,\bar{L}_0$ with $2$ constraints on them. In what follows, we shall present an explicit derivation for this. 

\subsection{Novel quadratic Casimirs and relevant tensor/spinor fields}
Partially inspired by \cite{MS,BKL,SS} and based on our experience acquired from \cite{Zhang,Zhang2}, we would like to define a novel quadratic Casimir operator
\begin{equation}
	\bar{C}^2=\bar{L}^2-\frac{1}{4}D^2
\end{equation}
associated with $sl(2,\mathbb{C})_R\times D$ Lie algebra.
Then we further introduce 
the following two auxiliary tensor fields
\begin{eqnarray}
	\bar{H}^{ab}&=&\bar{L}_0^a\bar{L}_0^b-\frac{1}{2}(\bar{L}_{-1}^a\bar{L}_{1}^b+\bar{L}_{1}^a\bar{L}_{-1}^b)-\frac{1}{4}D^aD^b,\nonumber\\
	\bar{Z}_{abc}&=&\bar{L}_{0a}\nabla_b \bar{L}_{0c}-\frac{1}{2}(\bar{L}_{-1a}\nabla_b \bar{L}_{1c}+\bar{L}_{1a}\nabla_b \bar{L}_{-1c})\nonumber\\
 &&-\frac{1}{4}{D}_a\nabla_b{D}_c.
\end{eqnarray}
Similarly, one can introduce the corresponding Casimir operator and associated auxiliary fields for the $sl(2,\mathbb{C})_L\times D$ Lie algebra, which are obviously related to the right ones by the complex conjugation. A straightforward calculation gives 
\begin{eqnarray}
\bar{H}^{ab}&=&-\frac{1}{4}x^2\eta^{ab},\nonumber\\
	\bar{Z}_{abc}&=&\frac{1}{4}(\eta_{ab}{D}_{c}-\eta_{bc}{D}_a-\eta_{ac} {D}_b-i\epsilon_{abcd}D^d).    
\end{eqnarray}
Whence we further have
\begin{eqnarray}
\bar{ Z}^a{}_{ac}&=&\frac{1}{2}D_c,\nonumber\\
\nabla_d\bar{Z}_{abc}&=&\frac{1}{4}(\eta_{ab}\eta_{cd}-\eta_{bc}\eta_{ad}-\eta_{ac}\eta_{bd}-i\epsilon_{abcd}),    
\end{eqnarray}
where we have used the fact $\nabla_aD_b=\eta_{ab}$. For the convenience of the later spinor analysis, we also like to introduce some spinor fields associated with our novel Casimirs as follows
\begin{widetext}
\begin{eqnarray}
	\alpha_{AA'B}{}^C(R)&=& \bar{L}_{0AA'}\Gamma_B{}^C(\bar{L}_0)-\frac{1}{2}(\bar{L}_{-1AA'}\Gamma_B{}^C(\bar{L}_1)+\bar{L}_{1AA'}\Gamma_B{}^C(\bar{L}_{-1}))-\frac{1}{4}{D}_{AA'}\Gamma_B{}^C(D)\nonumber\\
	&=&\frac{1}{2}\bar{Z}_{AA'BB'}{}^{CB'}+\frac{1}{8}D_{AA'}\delta_B{}^C=-\frac{1}{4}D_{AA'}\delta_B{}^C+\frac{1}{8}D_{AA'}\delta_B{}^C=-\frac{1}{8}D_{AA'}\delta_B{}^C,\nonumber\\
	\alpha_{AA'B}{}^C(L)&=& L_{0AA'}\Gamma_B{}^C(L_0)-\frac{1}{2}(L_{-1AA'}\Gamma_B{}^C(L_1)+L_{1AA'}\Gamma_B{}^C(L_{-1}))-\frac{1}{4}{D}_{AA'}\Gamma_B{}^C(D)\nonumber\\
	&=&\frac{1}{2}Z_{AA'BB'}{}^{CB'}+\frac{1}{8}D_{AA'}\delta_B{}^C\nonumber\\
	&=&-\frac{1}{4}(\epsilon_{AB}D^C{}_{A'}+\epsilon_A{}^CD_{BA'}+D_{AA'}\delta_B{}^C)+\frac{1}{8}D_{AA'}\delta_B{}^C\nonumber\\
	&=&-\frac{1}{4}(\epsilon_{AB}D^C{}_{A'}+\epsilon_A{}^CD_{BA'})-\frac{1}{8}D_{AA'}\delta_B{}^C,
\end{eqnarray}
\end{widetext}
and
\begin{widetext}
\begin{eqnarray}
	&&\gamma_{AD}{}^{BC}(R)=\nonumber\\
	&&\Gamma_A{}^B(\bar{L}_0)\Gamma_D{}^C(\bar{L}_0)-\frac{1}{2}(\Gamma_A{}^B(\bar{L}_{-1})\Gamma_D{}^C(\bar{L}_1)+\Gamma_A{}^B(\bar{L}_1)\Gamma_D{}^C(\bar{L}_{-1}))-\frac{1}{4}\Gamma_A{}^B(D)\Gamma_D{}^C(D)\nonumber\\
	&&=\frac{1}{4}(\nabla_{DD'}\bar{Z}^{CD'}{}_{AB'}{}^{BB'}+\frac{3}{4}\delta_A{}^B\delta_D{}^C)=\frac{1}{4}(-\delta_A{}^B\delta_D{}^C+\frac{3}{4}\delta_A{}^B\delta_D{}^C)=-\frac{1}{16}\delta_A{}^B\delta_D{}^C,\nonumber\\
	&&\gamma_{AD}{}^{BC}(L)=\nonumber\\
	&&\Gamma_A{}^B(L_0)\Gamma_D{}^C(L_0)-\frac{1}{2}(\Gamma_A{}^B(L_{-1})\Gamma_D{}^C(L_1)+\Gamma_A{}^B(L_1)\Gamma_D{}^C(L_{-1}))-\frac{1}{4}\Gamma_A{}^B(D)\Gamma_D{}^C(D)\nonumber\\
	&&=\frac{1}{4}(\nabla_{DD'}Z^{CD'}{}_{AB'}{}^{BB'}+\frac{3}{4}\delta_A{}^B\delta_D{}^C)=\frac{1}{4}(\delta_A{}^C\delta_D{}^B+\epsilon_{AD}\epsilon^{CB}-\delta_A{}^B\delta_D{}^C+\frac{3}{4}\delta_A{}^B\delta_D{}^C)\nonumber\\
	&&=\frac{1}{4}\delta_A{}^C\delta_D{}^B+\frac{1}{4}\epsilon_{AD}\epsilon^{CB}-\frac{1}{16}\delta_A{}^B\delta_D{}^C,
\end{eqnarray}
\end{widetext}
where we have used the following identities
\begin{eqnarray}
	\eta_{AA'BB'}&=&\epsilon_{AB}\epsilon_{A'B'}, \nonumber\\
 \epsilon_{AA'BB'CC'DD'}&=&i(\epsilon_{AB}\epsilon_{CD}\epsilon_{A'C'}\epsilon_{B'D'}\nonumber\\
 &&-\epsilon_{A'B'}\epsilon_{C'D'}\epsilon_{AC}\epsilon_{BD})
 \end{eqnarray}
with $\Gamma$ defined later on in Eq. (\ref{dLie}). Whence we further have
\begin{eqnarray}
	\beta_C{}^A(R)\equiv\gamma_{CD}{}^{DA}(R)&=&-\frac{1}{16}\delta_C{}^A,\nonumber\\
 \beta_C{}^A(L)\equiv\gamma_{CD}{}^{DA}(L)&=&\frac{11}{16}\delta_C{}^A.
\end{eqnarray}

\subsection{Two constraints on $sl(2,\mathbb{C})$ Casimirs and dilatation}
As a warm-up, let us start with the massless scalar field $\phi$, whose
equation of motion is given by
\begin{equation}
	\nabla_a\nabla^a\phi=0.
\end{equation}
The Lie derivative acting on the scalar field yields
\begin{eqnarray}
	\mathcal{L}_X\mathcal{L}_Y\phi&=&X^a\nabla_a(Y^b\nabla_b\phi)\nonumber\\
& =&(X^a\nabla_aY^b)\nabla_b\phi+X^aY^b\nabla_a\nabla_b\phi,
\end{eqnarray}
whereby we obtain
\begin{equation}
	\bar{\mathcal{C}}^2\phi=\bar{Z}^a{}_a{}^b\nabla_b\phi+\bar{H}^{ab}\nabla_a\nabla_b\phi=\frac{1}{2}D^a\nabla_a\phi=\frac{1}{2}\mathcal{D}\phi
\end{equation}
with $\mathcal{D}\equiv\mathcal{L}_D$.
Thus we have
\begin{equation}
	\mathcal{\bar{L}}^2\phi=(\frac{1}{4}\mathcal{D}^2+\frac{1}{2}\mathcal{D})\phi.
\end{equation}
It is easy to see that we also have
\begin{equation}
	\mathcal{L}^2\phi=(\frac{1}{4}\mathcal{D}^2+\frac{1}{2}\mathcal{D})\phi.
\end{equation}

Now let us move onto the massless spinor field with the helicity $s=-\frac{1}{2}$, whose dynamics is governed by the Weyl equation\footnote{Kindly please refer to Appendix \ref{C} for the relation between the helicity and the spinor index.}
\begin{equation}
	\nabla_{A'A}\phi^A=0.
\end{equation}
Not only is this equation equivalent to $\nabla_{A'}{}^A\phi^B=\nabla_{A'}{}^{(A}\phi^{B)}$, but also implies $\nabla_a\nabla^a\phi^A=0$.
To proceed, let us first recall the definition of Lie derivative of the spinor field $\phi^A{}_B$ with respect to a conformal Killing vector field $\xi^a$\cite{CD}
\begin{equation}
	\mathcal{L}_\xi\phi^A{}_B=\xi^a\nabla_a\phi^A{}_B-\phi^C{}_B\Gamma_C{}^A(\xi)+\phi^A{}_C\Gamma_B{}^C(\xi)
\end{equation}
with 
\begin{equation}\label{dLie}
	\Gamma_B{}^A(\xi)=\frac{1}{2}(\nabla_{BB'}\xi^{AB'}-\frac{1}{4}\nabla_c\xi^c\delta_B{}^A).
\end{equation}
The generalization of this definition to other types of spinor fields is obvious. It is noteworthy that $\nabla_a\Gamma_B{}^A(\xi)=0$ when restricted to the dilatation and Killing vector fields in our Minkowski spacetime. Thus with $X$ and $Y$ such vector fields, we have
\begin{eqnarray}
	\mathcal{L}_X\mathcal{L}_Y\phi^A&=&X^a\nabla_a(\mathcal{L}_Y\phi^A)-\mathcal{L}_Y\phi^B\Gamma_B{}^A(X)\nonumber\\
	&=&X^a\nabla_a(Y^b\nabla_b\phi^A-\phi^B\Gamma_B{}^A(Y))\nonumber\\
& &-(Y^b\nabla_b\phi^B-\phi^C\Gamma_C{}^B(Y))\Gamma_B{}^A(X)\nonumber\\
&	=&(X^a\nabla_aY^b)\nabla_b\phi^A+X^aY^b\nabla_a\nabla_b\phi^A\nonumber\\
&	&-\nabla_a\phi^B(X^a\Gamma_B{}^A(Y)+Y^a\Gamma_B{}^A(X))\nonumber\\
& &+\phi^C\Gamma_C{}^B(Y)\Gamma_B{}^A(X).
\end{eqnarray}
Whence we have
\begin{eqnarray}
    	\mathcal{\bar{C}}^2\phi^A&=&\bar{Z}^a{}_a{}^b\nabla_b\phi^A+\bar{H}^{ab}\nabla_a\nabla_b\phi^A\nonumber\\
 &    &-2\alpha^{CC'}{}_B{}^A(R)\nabla_{CC'}\phi^B+\phi^C\beta_C{}^A(R) \nonumber\\
&	=&\frac{1}{2}D^b\nabla_b\phi^A+\frac{1}{4}D^c\nabla_c\phi^A-\frac{1}{16}\phi^A\nonumber\\
& =&\frac{3}{4}\mathcal{D}\phi^A+\frac{5}{16}\phi^A,
\end{eqnarray}
where we have used $\mathcal{D}\phi^A=D^a\nabla_a\phi^A-\frac{1}{2}\phi^A$ in the last step. Similarly, we also have
\begin{eqnarray}
	\mathcal{C}^2\phi^A&=&\frac{1}{2}D^b\nabla_b\phi^A+\frac{1}{4}D^c\nabla_c\phi^A+\frac{11}{16}\phi^A\nonumber\\
 &&+\frac{1}{2}(\epsilon^C{}_BD^{AC'}+\epsilon^{CA}D_B{}^{C'})\nabla_{CC'}\phi^B\nonumber\\
&=&\frac{3}{4}D^b\nabla_b\phi^A+\frac{1}{2}D_B{}^{C'}\nabla_{C'}{}^A\phi^B+\frac{11}{16}\phi^A\nonumber\\
&=&\frac{3}{4}D^b\nabla_b\phi^A+\frac{1}{2}D^b\nabla_b\phi^A+\frac{11}{16}\phi^A\nonumber\nonumber\\
&	=&\frac{5}{4}\mathcal{D}\phi^A+\frac{21}{16}\phi^A.
\end{eqnarray}
Thus we end up with the following result
\begin{eqnarray}
	\mathcal{\bar{L}}^2\phi^A&=&(\frac{1}{4}\mathcal{D}^2+\frac{3}{4}\mathcal{D}+\frac{5}{16})\phi^A,\nonumber\\
 \mathcal{L}^2\phi^A&=&(\frac{1}{4}\mathcal{D}^2+\frac{5}{4}\mathcal{D}+\frac{21}{16})\phi^A.
 \end{eqnarray}

Next let us consider the Maxwell equation 
\begin{equation}
	\nabla_{A'A}\phi^{AB}=0
\end{equation}
for the electromagnetic field with $s=-1$, where $\phi^{AB}$ is symmetric with respect to $A$ and $B$. The Lie derivative acting on $\phi^{AB}$ gives rise to 
\begin{eqnarray}
	\mathcal{L}_X\mathcal{L}_Y\phi^{AB}&=&X^a\nabla_a(\mathcal{L}_Y\phi^{AB})-2\mathcal{L}_Y\phi^{C(B}\Gamma_C{}^{A)}(X)\nonumber\\
	&=&X^a\nabla_a(Y^b\nabla_b\phi^{AB}-2\phi^{C(B}\Gamma_C{}^{A)}(Y))\nonumber\\
 &&-2Y^b\nabla_b\phi^{C(B}\Gamma_C{}^{A)}(X)\nonumber\\
&	&+2\phi^{D(B}\Gamma_D{}^{|C|}(Y)\Gamma_C{}^{A)}(X)\nonumber\\
& &+2\phi^{CD}\Gamma_D{}^{(B}(Y)\Gamma_C{}^{A)}(X)\nonumber\\
	&=&(X^a\nabla_aY^b)\nabla_b\phi^{AB}+X^aY^b\nabla_{a}\nabla_b\phi^{AB}\nonumber\\
	&&-2\nabla_a\phi^{C(B}(X^a\Gamma_C{}^{A)}(Y)+Y^a\Gamma_C{}^{A)}(X))\nonumber\\
	&&+2\phi^{D(B}\Gamma_D{}^{|C|}(Y)\Gamma_C{}^{A)}(X)\nonumber\\
 &&+2\phi^{CD}\Gamma_D{}^{(B}(Y)\Gamma_C{}^{A)}(X),
 \end{eqnarray}
whereby we have
\begin{eqnarray}
	\mathcal{\bar{C}}^2\phi^{AB}&=&\frac{1}{2}D^a\nabla_a\phi^{AB}+\frac{1}{2}D^a\nabla_a\phi^{AB}-\frac{1}{8}\phi^{AB}-\frac{1}{8}\phi^{AB}\nonumber\\
	&=&D^a\nabla_a\phi^{AB}-\frac{1}{4}\phi^{AB}=\mathcal{D}\phi^{AB}+\frac{3}{4}\phi^{AB},\nonumber\\
	\mathcal{C}^2\phi^{AB}&=&\frac{1}{2}D^a\nabla_a\phi^{AB}+\frac{1}{2}D^a\nabla_a\phi^{AB}+D^a\nabla_a\phi^{AB}\nonumber \\
 &&+\frac{11}{8}\phi^{AB}+\frac{3}{8}\phi^{AB}=2D^a\nabla_a\phi^{AB}+\frac{7}{4}\phi^{AB}\nonumber\\
 &=&2\mathcal{D}\phi^{AB}+\frac{15}{4}\phi^{AB}.
\end{eqnarray}
Thus we wind up with 
\begin{eqnarray}
	\mathcal{\bar{L}}^2\phi^{AB}&=&(\frac{1}{4}\mathcal{D}^2+\mathcal{D}+\frac{3}{4})\phi^{AB},\nonumber\\
 \mathcal{L}^2\phi^{AB}&=&(\frac{1}{4}\mathcal{D}^2+2\mathcal{D}+\frac{15}{4})\phi^{AB}.
 \end{eqnarray}
 
The above spinor analysis has already involved all the necessary ingredients for one to obtain the corresponding result for the massless spinor field with other helicities. To be more specific, one can find
\begin{eqnarray}\label{minus}
	\mathcal{\bar{L}}^2\phi^{A\cdots}&=&(\frac{1}{4}\mathcal{D}^2+\frac{n+2}{4}\mathcal{D}+\frac{n^2+4n}{16})\phi^{A\cdots}\nonumber 
 \\&=&(\frac{\mathcal{D}}{2}-\frac{s}{2}+1)(\frac{\mathcal{D}}{2}-\frac{s}{2})\phi^{A\cdots},\nonumber\\
	\mathcal{L}^2\phi^{A\cdots}&=&(\frac{1}{4}\mathcal{D}^2+\frac{3n+2}{4}\mathcal{D}+\frac{9n^2+12n}{16})\phi^{A\cdots}\nonumber\\
 &=&(\frac{\mathcal{D}}{2}-\frac{3s}{2}+1)(\frac{\mathcal{D}}{2}-\frac{3s}{2})\phi^{A\cdots}
\end{eqnarray}
for the massless spinor field with helicity $s=-\frac{n}{2}$, which is totally symmetric with respect to $n$ indices and satisfies the equation of motion
\begin{equation}
	\nabla_{A'A}\phi^{A\cdots}=0.
\end{equation}
Note that the massless spinor field $\phi^{A'B'C'\cdots}$ with helicity $s=\frac{n}{2}$ is simply the complex conjugation of the massless spinor field with helicity $s=-\frac{n}{2}$, so we have 
\begin{eqnarray}\label{plus}
	\mathcal{\bar{L}}^2\phi^{A'\cdots}&=&(\frac{1}{4}\mathcal{D}^2+\frac{3n+2}{4}\mathcal{D}+\frac{9n^2+12n}{16})\phi^{A'\cdots}\nonumber\\
 &=&(\frac{\mathcal{D}}{2}+\frac{3s}{2}+1)(\frac{\mathcal{D}}{2}+\frac{3s}{2})\phi^{A'\cdots},\nonumber\\
	\mathcal{L}^2\phi^{A'\cdots}&=&(\frac{1}{4}\mathcal{D}^2+\frac{n+2}{4}\mathcal{D}+\frac{n^2+4n}{16})\phi^{A'\cdots}\nonumber\\
 &=&(\frac{\mathcal{D}}{2}+\frac{s}{2}+1)(\frac{\mathcal{D}}{2}+\frac{s}{2})\phi^{A'\cdots}
\end{eqnarray}
for the massless spinor field with helicity $s=\frac{n}{2}$.

On the  other hand,  by Eq. (\ref{reptrans}) for a conformal primary wavefunction, we have 
\begin{equation}\label{weight}
	\mathcal{L}^2=h(h-1),\quad \bar{\mathcal{L}}^2=\bar{h}
	(\bar{h}-1).
\end{equation}
So it is reasonable to expect that the candidate basis out of the simultaneous eigenvectors of $D, L^2, L_0,\bar{L}^2,\bar{L}_0$ can be constructed in terms of the infinite tower of descendants of the left and right highest (lowest) weight conformal primary wavefunction of $sl(2,\mathbb{C})$ Lie algebra, where the celestial conformal weights are determined by its bulk scaling dimension\footnote{The basis constructed in this way is discrete, compared to the frequently considered one from the unitary principal series.}. Actually, it has been shown in \cite{CMS} that this is the case for the massless scalar field. Eq. (\ref{minus}) and Eq. (\ref{plus}) obtained here provide us with an important foundation to generalize \cite{CMS} to the massless field with arbitrary helicity. In the subsequent section, we shall specify the explicit correspondence between the 2D celestial conformal weights and the 4D bulk scaling dimension for all the on-shell conformal primary wavefunctions. 
\section{Correspondence between the 4D bulk scaling dimension and 2D celestial conformal weights}\label{correspondence}
By Eq.  (\ref{weight}) and Eq. (\ref{minus}), we have the following relationship between the celestial conformal weight and bulk scaling dimension
\begin{equation}
	R_+:   h=\frac{\mathcal{D}}{2}-\frac{3s}{2}+1, \quad \texttt{or}\quad R_-: h=-\frac{\mathcal{D}}{2}+\frac{3s}{2},
\end{equation}
and
\begin{equation}
	\bar{R}_+: \bar{h}=\frac{\mathcal{D}}{2}-\frac{s}{2}+1, \quad \texttt{or}\quad \bar{R}_-: \bar{h}=-\frac{\mathcal{D}}{2}+\frac{s}{2}
\end{equation}
for the on-shell conformal primary wavefunctions with negative helicity $s$. However, this relationship demonstrates a certain ambiguity. To fix it, we like to find the explicit expression for the on-shell conformal primary wavefunctions.  As such, we follow \cite{PP,PPP} to choose $o^A=\frac{1}{\sqrt{q\cdot x}}\lambda^A, \iota^A=D^{AA'}o_{A'}$. Accordingly, we have
\begin{equation}
	\nabla_{A'A}o^B=-\frac{1}{2}o_{A'}o_Ao^B,\quad \nabla_{A'A}\iota^B=\delta_A{}^Bo_{A'}-\frac{1}{2}o_{A'}o_A\iota^B,
\end{equation}
which implies 
\begin{equation}
	\nabla_{A'A}o^A=0,\quad \nabla_{A'A}\iota^A=\frac{3}{2}o_{A'}.
\end{equation}
Whence we further have 
\begin{widetext}
\begin{eqnarray}
	\nabla_{A'A}o^{(AB\cdots}\iota^{CD\cdots)}\nonumber&=&\frac{m}{m+n}\nabla_{A'A}(o^Ao^{(B\cdots}\iota^{CD\cdots)}+\frac{n}{m+n}\nabla_{A'A}o^{(CD\cdots}\iota^{|A|}\iota^{B\cdots)}\nonumber\\
	&=&\frac{mn}{m+n}o^{(BC\cdots}\iota^{D\cdots)}o_{A'}\nonumber-\frac{mn-3n-n(n-1)}{2(m+n)}o^{(CD\cdots}\iota^{B\cdots)}o_{A'}\nonumber\\
	&=&\frac{n(n+m+2)}{2(m+n)}o^{(BC\cdots}\iota^{D\cdots)}o_{A'},
\end{eqnarray}    
\end{widetext}
where $o^{AB\cdots}$ denotes the spinor field produced by the product of $m$ $o$s and $\iota^{CD\cdots}$ denotes the spinor field produced by the product of $n$ $\iota$s. With this, we can construct the following linearly independent on-shell conformal primary wavefunctions for the massless spinor field with negative helicity $s$
\begin{eqnarray}
 	\phi^{AB\cdots}&=&\phi^\Delta(x)o^{AB\cdots}, \nonumber\\ 
  \hat{\phi}^{AB\cdots}&=&\phi^{-s+1}(x)o^{(A\cdots}\iota^{B\cdots)},\nonumber\\
  \tilde{\phi}^{AB\cdots}&=&(x^2)^{\Delta+s-1}\phi^\Delta(x)\iota^{AB\cdots},   
\end{eqnarray}
where we have used $D_{AA'}D^{AB'}=\frac{1}{2}x^2\delta_{A'}{}^{B'}$ with the scalar function defined as $\phi^\Delta(x)=\frac{1}{(q\cdot x)^\Delta}$. It is noteworthy that besides the first and third kinds of on-shell conformal primary wavefunctions, which are familiar to the community and related to each other by the so-called shadow transformation, we also find the second kind of on-shell conformal primary wavefunctions for $s\le -1$. According to Table.\ref{list1}, we further list the celestial conformal weights and bulk scaling dimension for the above explicit on-shell  conformal primary wavefunctions in Table \ref{list2}\footnote{It is noteworthy that the second kind of on-shell conformal primary wavefunctions displays a different correspondence between the celestial spin and bulk helicity from the first and third ones, whose celestial spin is related to the bulk helicity simply by $J=\pm s$.}. Whence we obtain a definite relationship between the celestial conformal weights and bulk scaling dimension for each on-shell conformal primary wavefunction, i.e., 
\begin{table*}
	\centering
	\begin{tabular}{c|c|c|c|c|c}
		& $h$ & $\bar{h}$& $\Delta$ & $J$ & $\mathcal{D}$\\\hline
		$\phi^{AB\cdots}$  & $\frac{\Delta+s}{2}$& $\frac{\Delta-s}{2}$& $\Delta $&$s$ & $-\Delta +2s$\\
		$\hat{\phi}^{AB\cdots}$& $-s-\frac{o}{2}+\frac{1}{2}$ & $\frac{o}{2}+\frac{1}{2}$&$-s+1$ & $-s-o$ & $s-o-1$\\
		$\tilde{\phi}^{AB\cdots}$& $\frac{\Delta-s}{2}$& $\frac{\Delta+s}{2}$   &$\Delta$& $-s$ & $\Delta+2s-2$\\
	\end{tabular}
	\caption{The 2D celestial conformal weights (conformal dimension and spin) and 4D bulk scaling dimension for the on-shell conformal primary wavefunctions with negative helicity $s$, where $o$ denotes the number of $o^A$ in $\hat{\phi}^{AB\cdots}$.}
	\label{list2}
\end{table*}
\begin{eqnarray}
	R_- \quad \texttt{and} \quad \bar{R}_- \quad && \texttt{ for} \quad \phi^{AB\cdots}, \nonumber\\
	R_+ \quad \texttt{and} \quad \bar{R}_- \quad && \texttt{for}\quad \hat{\phi}^{AB\cdots}, \nonumber\\
	R_+ \quad \texttt{and} \quad \bar{R}_+ \quad &&\texttt{for} \quad \tilde{\phi}^{AB\cdots}.
\end{eqnarray}

As pointed out before, the on-shell massless spinor field with positive helicity is simply the complex conjugation of that with negative helicity. So it is not hard to obtain the parallel results for the massless spinor field with positive helicity, which we shall present below for completeness. Namely, Eq. (\ref{weight}) together with Eq. (\ref{plus}) gives rise to
\begin{equation}
	R_+:  h=\frac{\mathcal{D}}{2}+\frac{s}{2}+1, \quad \texttt{or}\quad R_-: h=-\frac{\mathcal{D}}{2}-\frac{s}{2},
\end{equation}
and 
\begin{equation}
	\bar{ R}_+:  \bar{ h}=\frac{\mathcal{D}}{2}+\frac{3s}{2}+1, \quad \texttt{or}\quad \bar{R}_-: \bar{ h}=-\frac{\mathcal{D}}{2}-\frac{3s}{2}.
\end{equation}
Such an ambiguity in the correspondence between the celestial conformal weights and bulk scaling dimension can be resolved by examining the explicit quantities for each kind of on-shell conformal primary wavefunctions in Table \ref{list3}. As a result, we have 
\begin{eqnarray}
	&& R_- \quad \texttt{and} \quad \bar{R}_- \quad  \texttt{ for} \quad \phi^{A'B'\cdots}, \nonumber\\
	&&  R_- \quad \texttt{and} \quad \bar{R}_+ \quad \texttt{for}\quad \hat{\phi}^{A'B'\cdots}, \nonumber\\
	&&  R_+ \quad \texttt{and} \quad \bar{R}_+ \quad \texttt{for} \quad \tilde{\phi}^{A'B'\cdots}.
\end{eqnarray}

\begin{table*}
	\begin{tabular}{c|c|c|c|c|c}
		& $h$ & $\bar{h}$& $\Delta$ & $J$ & $\mathcal{D}$\\ \hline
		$\phi^{A'B'\cdots}$  & $\frac{\Delta+s}{2}$& $\frac{\Delta-s}{2}$& $\Delta $&$s$ & $-\Delta -2s$\\
		$\hat{\phi}^{A'B'\cdots}$ & $\frac{o}{2}+\frac{1}{2}$& $s-\frac{o}{2}+\frac{1}{2}$
		&$s+1$ & $-s+o$ & $-s-o-1$\\
		$\tilde{\phi}^{A'B'\cdots}$& $\frac{\Delta-s}{2}$& $\frac{\Delta+s}{2}$   &$\Delta$& $-s$ & $\Delta-2s-2$\\
	\end{tabular}
	\caption{The 2D celestial conformal weights (conformal dimension and spin) and 4D bulk scaling dimension for the on-shell conformal primary wavefunctions  with positive helicity $s$, obtained by taking the complex conjugation of those with negative helicity $-s$, where $o$ denotes the number of $o^{A'}$ in $\hat{\phi}^{A'B'\cdots}$.}
	\label{list3}
\end{table*}

\section{Discussion}
Although the bulk dilatation does not belong to the isometry Poincar\'e group of our Minkowski spacetime, it can be regarded as an emergent symmetry of the solution space of equations of motion for the massless field dictated by the unitary representation of the Poincar\'e group, reminiscent of the hidden conformal symmetry of the Kerr black hole discovered in \cite{CMS1}. With this in mind, we have shown that the $sl(2,\mathbb{C})\times D$ symmetry dictated candidate basis for the massless particle representation of the Poincar\'e group can be constructed out of the infinite tower of the descendants of the left and right highest (lowest) weight conformal primary wavefunction of $sl(2,\mathbb{C})$ Lie algebra, where the celestial conformal weights are further determined in an explicit manner by the bulk scaling dimension through solving out the exact on-shell conformal primary wavefunctions for the massless field with arbitrary helicity. In particular, on top of the two kinds of familiar-looking on-shell conformal primary wavefunctions, which are related to each other by the shadow transformation,  we also find another set of independent on-shell conformal primary wavefunctions for the massless field with helicity $|s|\ge 1$. In addition, for the massless field with helicity $|s|\ge 1$, one is also required to introduce the gauge potential to define the Klein-Gordon inner product\cite{Penrose}. So we present the exact on-shell conformal primary wavefunctions as well as the corresponding celestial conformal weights and bulk scaling dimension in Appendix \ref{D} for the electromagnetic potential, which is supposed to be generalized readily to the massless field with larger helicity. 

However, to show that our candidate basis is really the basis for the massless particle representation of the Poincar\'e group, one is required to show that it is complete as a basis, which has already been analyzed in \cite{CMS,FPR,Mitra}, where different strategies are developed. It is interesting to ask whether one can achieve its completeness in another manner, where the aforementioned new set of independent on-shell conformal primary wavefunctions may be an indispensable part. We hope to address this important issue elsewhere in the future.

\section*{Acknowledgements}
We are grateful to Yichen Feng, Shengyi Liu, Sirui Shuai, Yu Tian, and Xiaoning Wu for their stimulating discussions. This work is partly supported by the National Key Research and Development Program of China with Grant No. 2021YFC2203001 as well as the National Natural Science Foundation of China with Grant Nos. 12075026 and 12361141825.

\appendix

\section{ Conformal algebra in the $d$-dimensional Minkowski spacetime}\label{A}
For $d$-dimensional Minkowski spacetime with $x^\mu$ the Lorentz coordinates, the global conformal Killing vector fields can be written as
\begin{eqnarray}
&&P_\mu=\partial_\mu,\quad D=x^\mu\partial_\mu,\nonumber\\
&&M_{\mu\nu}=x_\mu\partial_\nu-x_\nu\partial_\mu, \nonumber\\
&&K_\mu=2x_\mu x^\nu\partial_\nu-x^2\partial_\mu
 \end{eqnarray}
with the non-vanishing commutation relations given by
\begin{eqnarray}
	&&[D,P_\mu]=-P_\mu,\quad [P_\rho,M_{\mu\nu}]=\eta_{\rho\mu}P_\nu-\eta_{\rho\nu}P_\mu,\nonumber\\
	&& [M_{\mu\nu},M_{\rho\sigma}]=-(\eta_{\mu\rho}M_{\nu\sigma}-\eta_{\mu\sigma}M_{\nu\rho}-\eta_{\nu\rho}M_{\mu\sigma}+\eta_{\nu\sigma}M_{\mu\rho}),\nonumber\\
	&&[D, K_\mu]=K_\mu,\quad [K_\mu,P_\nu]=-2(\eta_{\mu\nu}D+M_{\mu\nu}).
\end{eqnarray}
\section{The formula for the complex coordinate with $w=0$ as the north pole}\label{B}
Note that the north pole itself corresponds to $w=\infty$ in the complex coordinate $w$ given by the north pole based stereographic projection, so to circumvent the potential subtleties associated with the north pole, we prefer to choose $\lambda^\Sigma=(1,w)$. Accordingly, $q^\mu=(1+w\bar{w},w+\bar{w},i(w-\bar{w}),1-w\bar{w})$ with the north pole located at $w=0$. As a result, Eq. (\ref{coordinatetrans}) and Eq. (\ref{3}) will be modified as follows
\begin{equation}
	w'=\frac{c+dw}{a+bw},\quad \bar{w}'=\frac{\bar{c}+\bar{d}\bar{w}}{\bar{a}+\bar{b}\bar{w}},
\end{equation}
and
\begin{eqnarray}
	l_{-1}\rightarrow -T_1,\quad && \bar{l}_{-1}\rightarrow -\bar{T}_1,\nonumber\\
	l_1\rightarrow -T_{-1},\quad &&\bar{l}_1\rightarrow -\bar{T}_{-1},\nonumber\\
	l_0\rightarrow -T_0,\quad && \bar{l}_0\rightarrow -\bar{T}_0.
\end{eqnarray}
Similarly,  Eq. (\ref{celestial}) and Eq. (\ref{reptrans}) will also get modified in the following way
\begin{eqnarray}
&&	\mathcal{O}(x'^\mu=\Lambda^\mu{}_\nu x^\nu;w'=\frac{c+dw}{a+bw},\bar{w}'=\frac{\bar{c}+\bar{d}\bar{w}}{\bar{a}+\bar{b}\bar{w}})\nonumber\\
&& =|\frac{\partial w'}{\partial w}|^{-\frac{\Delta+J}{2}}|\frac{\partial\bar{w}'}{\partial\bar{w}}|^{-\frac{\Delta-J}{2}}D(\Lambda)\mathcal{O}(x;w,\bar{w})
\end{eqnarray}
with $|\frac{\partial w'}{\partial w}|=\frac{1}{(a+bw)^2}$ and 
\begin{eqnarray}
	&& \mathcal{L}_{L_n}\mathcal{O}=\mathcal{L}_{T_{-n}}\mathcal{O}=(w^{1-n}\partial_w+h(1-n)w^{-n})\mathcal{O},\nonumber\\
	&&\mathcal{L}_{\bar{L}_n}\mathcal{O}=\mathcal{L}_{\bar{T}_{-n}}\mathcal{O}=(\bar{w}^{1-n}\partial_{\bar{w}}+\bar{h}(1-n)\bar{w}^{-n})\mathcal{O}.\nonumber\\
\end{eqnarray}
\section{$s=-\frac{n}{2}$ for unprimed spinor fields and $s=\frac{n}{2}$ for primed spinor fields}\label{C}
\begin{table*}[ht]
	\centering
	\begin{tabular}{c|c|c|c|c|c}
		& $h$ & $\bar{h}$& $\Delta$ & $J$ & $\mathcal{D}$\\\hline
		$\bar{m}_a\phi^\Delta(x) (\Delta=1, \texttt{gauge mode})$  & $\frac{\Delta-1}{2}$& $\frac{\Delta+1}{2}$& $\Delta $&$-1$ & $-\Delta+1 $\\
		$m_a(x^2)^{\Delta-1}\phi^\Delta(x) (\Delta=1, \texttt{gauge mode})$ & $\frac{\Delta+1}{2}$& $\frac{\Delta-1}{2}$
		&$\Delta$ & $1$ & $\Delta-1$\\
		$\bar{m}_a\phi^1(x)\ln\frac{ q\cdot x}{ q\cdot x_0}$& $0$& $1$   &$1$& $-1$ & $0 (\texttt{up to a gauge})$\\
		$m_a\phi^1(x)\ln(\frac{q\cdot x_0 }{q\cdot x}x^2)$& $1$& $0$   &$1$& $1$ & $0 (\texttt{up to a gauge})$\\
		$(-\frac{x^2}{2}n_a+l_a)\phi^2(x) (\texttt{gauge mode})$& $1$& $1$   &$2$& $0$ & $0$\\
		$(-\frac{x^2}{2}n_a+l_a)\phi^2(x)\ln x^2+$& $$& $$   &$$& $$ & $0$\\
		$\phi^1(x)(m_a\frac{\bar{m}\cdot x_0}{q\cdot x_0}-\bar{m}_a\frac{m\cdot x_0}{q\cdot x_0})$& $1$& $1$   &$2$& $0$ & $(\texttt{up to a gauge})$\\
	\end{tabular}
	\caption{The 2D celestial conformal weights (conformal dimension and spin) and 4D bulk scaling dimension for the on-shell conformal primary wavefunctions of the electromagnetic potential with negative helicity $-1$. For the on-shell celetial conformal primary wavefunctions with positive helicity $1$, which is the complex conjugation of those with negative helicity $-1$,  the corresponding result can be obtained by simply taking $h\leftrightarrow \bar{h}$ and $J\rightarrow -J$.}
	\label{list4}
\end{table*}
Obviously, the massless scalar field has zero helicity. On the other hand, as stated in the main body of our paper, unprimed and primed massless spinor fields have negative and positive helicities, respectively. Here we take the massless spinor field with one index as an example to show that with our convention this is the case, i.e.,
\begin{widetext}
\begin{eqnarray}
	&&W^\mu\phi^E=-\frac{1}{2}\epsilon^{\mu\nu\rho\sigma}\mathcal{L}_{P_\nu}\mathcal{L}_{M_{\rho\sigma}}\phi^E=\frac{1}{2}\epsilon^{\mu\nu\rho\sigma}\partial_\nu\phi^F\nabla_{FF'}x_\rho(\frac{\partial}{\partial x^\sigma})^{EF'}\nonumber\\
	&&=\frac{i}{2}\sigma^\mu{}_{AA'}(\epsilon^{AB}\epsilon^{CD}\epsilon^{A'C'}\epsilon^{B'D'}-\epsilon^{A'B'}\epsilon^{C'D'}\epsilon^{AC}\epsilon^{BD})\nabla_{BB'}\phi^F\epsilon_{CF}\epsilon_{C'F'}\delta_{D}{}^E\delta_{D'}{}^{F'}\nonumber\\
	&&=\sigma^\mu{}_{AA'}(\frac{i}{2}\epsilon^{AB}\epsilon^{A'B'}\nabla_{BB'}\phi^E-i\epsilon^{A'B'}\epsilon^{EB}\nabla_{B'B}\phi^A)\nonumber\\
	&&=-\frac{i}{2}\eta^{\mu\nu}\partial_\nu\phi^E=-\frac{i}{2}\mathcal{L}_{P^\mu}\phi^E,
\end{eqnarray}    
\end{widetext}
which tells us that the unprimed $\phi^E$ and the primed $\phi^{E'}$  have helicity $-\frac{1}{2}$ and $\frac{1}{2}$, respectively. 
\section{On-shell conformal primary wavefunctions of electromagnetic potential}\label{D}
For the Killing vector fields or the dilatation field $X$ and $Y$ in Minkowski spacetime, we have
\begin{widetext}
 \begin{eqnarray}
	\mathcal{L}_X\mathcal{L}_YA_c&&=X^a\nabla_a(Y^b\nabla_bA_c+A_b\nabla_cY^b)+(Y^b\nabla_bA_a+A_b\nabla_aY^b)\nabla_cX^a\nonumber\\
	&&=X^a\nabla_aY^b\nabla_bA_c+X^aY^b\nabla_a\nabla_bA_c+X^a\nabla_cY^b\nabla_aA_b+Y^b\nabla_cX^a\nabla_bA_a+\nabla_a(Y^b\nabla_cX^a)A_b\nonumber\\
\end{eqnarray}   
\end{widetext}
for our electromagnetic potential $A_c$, where we have used the fact the second derivative of $X$ and $Y$ vanishes. With the gauge condition $\nabla_cA^c=D^cA_c=0$, we further have 
\begin{eqnarray}
	\mathcal{\bar{C}}^2A_c&=&\bar{Z}^a{}_a{}^b\nabla_bA_c+\bar{H}^{ab}\nabla_a\nabla_bA_c+2\bar{Z}^a{}_c{}^b\nabla_aA_b\nonumber\\
 &&+\nabla_a(\bar{Z}^b{}_c{}^a)A_b\nonumber\\
	&=&\frac{1}{2}D^b\nabla_bA_c-\frac{1}{4}x^2\Box A_c-\frac{1}{2}\mathcal{D}A_c\nonumber\\
 &&+\frac{i}{2}\nabla_aA_b\epsilon^{ab}{}_{cd}D^d +\frac{1}{2}A_c\nonumber\\
 &=&\mp\frac{1}{2}F_{cd}D^d=\pm\frac{1}{2}\mathcal{D}A_c
\end{eqnarray}
for the helicities $s=\pm1$, which correspond to
\begin{equation}
	\frac{1}{2}\epsilon_{abcd}F^{cd}=\pm iF_{ab}
\end{equation}
through the relations $F^{AA'BB'}=\phi^{AB}\epsilon^{A'B'}$ for $s=-1$ and  $F^{AA'BB'}=\phi^{A'B'}\epsilon^{AB}$ for $s=1$.
Likewise, we have
\begin{equation}
	\mathcal{C}^2A_c=\mp \frac{1}{2}\mathcal{D}A_c
\end{equation}
for $s=\pm 1$. Thus we end up with
\begin{equation}
	\mathcal{L}^2=\frac{\mathcal{D}}{2}(\frac{\mathcal{D}}{2}-s),\quad \mathcal{\bar{L}}^2=\frac{\mathcal{D}}{2}(\frac{\mathcal{D}}{2}+s),
\end{equation}
which implies the relationship between the bulk scaling dimension and celestial conformal weights with a certain ambiguity as follows
\begin{eqnarray}
	&&  R_+: h=\frac{\mathcal{D}}{2}-\frac{s}{2}+\frac{1}{2},\quad \texttt{or}\quad  R_-: h=-\frac{\mathcal{D}}{2}+\frac{s}{2}+\frac{1}{2},\nonumber\\
	&& \bar{R}_+: \bar{h}=\frac{\mathcal{D}}{2}+\frac{s}{2}+\frac{1}{2},\quad \texttt{or} \quad \bar{R}_-: \bar{h}=-\frac{\mathcal{D}}{2}-\frac{s}{2}+\frac{1}{2}\nonumber\\
\end{eqnarray}
for the on-shell conformal primary wavefunctions of the electromagnetic potential. 
Furthermore, with the aforementioned gauge condition as well as the choice of the null tetrad
\begin{eqnarray}
&& l_{AA'}=\iota_A\iota_{A'},\quad n_{AA'}=o_Ao_{A'},\nonumber\\
 &&m_{AA'}=\iota_Ao_{A'},\quad \bar{m}_{AA'}=o_A\iota_{A'},   
\end{eqnarray}
the on-shell conformal primary wavefunctions of the electromagnetic potential can be constructed as follows
\begin{equation}
	A_a=\bar{m}_a\phi^\Delta(x),\quad  \tilde{A}_a=m_a(x^2)^{\Delta-1}\phi^\Delta(x),
\end{equation}
which correspond to 
\begin{eqnarray}
    	\phi^{AB}&=&(1-\Delta)\phi^\Delta(x)o^Ao^B,\nonumber\\
     \tilde{\phi}^{AB}&=&2(1-\Delta)(x^2)^{\Delta-2}\phi^\Delta(x)\iota^A\iota^B
\end{eqnarray}
through the relation $\phi_{AB}=\nabla_{C'(A}A^{C'}{}_{B)}=\nabla_{C'A}A^{C'}{}_B$. Here we have used 
\begin{equation}
	x_a=\frac{x^2}{2}n_a+l_a,
\end{equation}
and 
\begin{equation}
	\nabla_a\bar{m}_b=-\bar{m}_an_b,\quad \nabla_an_b=-n_an_b,
\end{equation}
which implies $\nabla_a\bar{m}^a=0$ and $\Box m_a=0$. Obviously, the above on-shell conformal wavefunctions with $\Delta=1$ correspond to the pure gauge modes. For $\Delta=1$, one can instead construct the non-gauge modes for the electromagnetic potential as follows
\begin{equation}
	A_a=\bar{m}_a\phi^1(x)\ln\frac{ q\cdot x}{ q\cdot x_0},\quad \tilde{A}_a=m_a\phi^1(x)\ln(\frac{q\cdot x_0 }{q\cdot x}x^2)
\end{equation}
with $x_0=(1,0,0,1)$ the reference point in our Minkowski spacetime,
which give rise to 
\begin{equation}
	\phi^{AB}=\phi^1(x)o^Ao^B,\quad \tilde{\phi}^{AB}=-2(x^2)^{-1}\phi^1(x)\iota^A\iota^B,
\end{equation}
respectively.
Last,  by 
\begin{equation}
	\nabla_al_b=n_al_b-m_a\bar{m}_b-\bar{m}_am_b,
\end{equation}
one can show that 
\begin{eqnarray}
\hat{A}_a&=&(-\frac{x^2}{2}n_a+l_a)\phi^2(x), \nonumber\\
\hat{A}_a&=&(-\frac{x^2}{2}n_a+l_a)\phi^2(x)\ln x^2\nonumber\\
&&+\phi^1(x)(m_a\frac{\bar{m}\cdot x_0}{q\cdot x_0}-\bar{m}_a\frac{m\cdot x_0}{q\cdot x_0})
\end{eqnarray}
are also the on-shell conformal primary wavefunctions of the electromagnetic potential, corresponding to a pure gauge and 
\begin{equation}
	\hat{\phi}^{AB}=4\phi^2(x)o^{(A}\iota^{B)},
\end{equation}
respectively. 

So we have already succeeded in obtaining the on-shell conformal primary wavefunctions of the electromagnetic potential and its field strength for the negative helicity. The corresponding celestial conformal weights and bulk scaling dimension are listed in Table \ref{list4}.  The result for the positive helicity can be readily obtained  by noting that the on-shell conformal primary wavefunctions for the positive helicity is related to those for the negative helicity by the complex conjugation. Whence the correpondence between the celestial conformal weights and bulk scaling dimension can be specified as follows
\begin{eqnarray}
	&&R_- \quad \texttt{and} \quad \bar{R}_- \quad\texttt{for}\quad A_a (\bar{A}_a),\nonumber\\
	&& R_+ \quad \texttt{and} \quad \bar{R}_- \quad\texttt{for}\quad \hat{A}_a,\nonumber\\
	&& R_- \quad \texttt{and} \quad \bar{R}_+ \quad\texttt{for}\quad \bar{\hat{A}}_a,\nonumber\\
	&&R_+ \quad \texttt{and} \quad \bar{R}_+ \quad\texttt{for}\quad \tilde{A}_a(\bar{\tilde{A}}_a).
\end{eqnarray}
	
\end{document}